\documentstyle[PASJadd,epsf]{PASJ95}

\newcommand{\vol}{}
\newcommand{\no}{}
\newcommand{\lett}{}
\newcommand{\spage}{}
\newcommand{\rdate}{}
\newcommand{\adate}{}
\begin{document}
\title{Discovery and Photometric Observation of the Optical Counterpart in 
a Possible Galactic Halo X-ray Transient, XTE J1118+480}
\author{Makoto {\sc Uemura},$^1$ Taichi {\sc Kato},$^1$ 
Katsura {\sc Matsumoto},$^1$ Hitoshi {\sc Yamaoka},$^2$ 
Kesao {\sc Takamizawa},$^3$\\ Yasuo {\sc Sano},$^4$ Katsumi {\sc Haseda},$^5$ 
Lewis M. {\sc Cook},$^6$ Denis {\sc Buczynski},$^7$ Gianluca {\sc Masi},$^8$ \\
{\it $^1$ Department of Astronomy, Faculty of Science, Kyoto University, Sakyou-ku, Kyoto 606-8502}\\
{\it E-mail (MU): uemura@kusastro.kyoto-u.ac.jp}\\
{\it $^2$ Department of Physics, Faculty of Science, Kyushu University, Fukuoka 810}\\
{\it $^3$ Oohinata 65-1,  Saku-machi, Nagano, 384-0502}\\
{\it $^4$ VSOLJ, Nishi juni-jou minami 3-1-5, Nayoro, Hokkaido}\\
{\it $^5$ VSOLJ, Hujimidai 2-7-10, Toyohashi, Aichi 441-8135}\\
{\it $^6$ Center for Backyard Astrophysics (California), 1730 Helix Ct. Concord, California 94518, USA}\\
{\it $^7$ Conder Brow Observatory, Littlefell Lane, Lancaster LA20RQ, England}\\
{\it $^8$ Center for Backyard Astrophysics (Italy), Via Madonna de Loco, 47, 03023 Ceccano, Italy}}
\abst{We discovered the optical counterpart about 13 mag of a soft X-ray 
transient, XTE J1118+480 on 2000 March 30.  We perform astrometry and 
provide the accurate position as R.A. = $11^{\rm h}\,18^{\rm m}\,10^{\rm s}.85$
, Decl. = $+48^\circ\,02^\prime\,12^{\prime \prime}.9$.  The outbursting 
object is identified with a 18.8 mag star in USNO catalog.  Our pre-discovery 
data shows another outburst during 2000 January, again coinciding with an 
outburst detected in X-rays.  Through 
the CCD time-series photometry, we found the presence of a periodic 
variation with the amplitude of 0.055 mag and the period of $0.17078 \pm 
0.00004\, {\rm d}$ which we consider as promising candidate of orbital 
period.  Because of the high galactic latitude and faint quiescence 
magnitude of 18.8, XTE J1118+480 is the possible first firmly identified 
black hole candidate (BHC) X-ray transient in the galactic halo.}
\kword{accretion, accretion disks --- stars: activity --- 
stars: individual (XTE J1118+480)}
\maketitle
\thispagestyle{headings}
\section{Introduction}
Soft X-ray transients (SXTs), or X-ray novae (e.g. Chen et al. 1997) 
are binary systems which exhibit luminous X-ray and optical outburst 
(Tanaka, Shibazaki 1996).  They uniquely provide the most compelling 
evidence for the existence of steller mass black holes using radial 
velocity studies, giving mass functions exceeding the maximum mass 
of a stable neutron star ($\sim 3\,M_\odot$), and we know eight such black 
hole candidates (van Paradijs, McClintock 1995; Bailyn et al. 1998; 
Orosz et al. 1998).  Their outburst light curves often have an common 
feature, that is, after a rise of a few days, the X-ray intensity comes 
to the maximum, which was typically followed by the exponential decay 
with an e-folding time of $\sim 40\,{\rm d}$ (Chen et al. 1997).  
At the maximum, the X-ray luminosity reaches $10^{38-39}\;{\rm erg\,s^{-1}}$ 
and the optical to X-ray flux ratio of $\sim 500$ is typical in SXTs 
(Tanaka, Shibazaki 1996).  

Many astronomers have recently believed that 
the compact object of SXTs in quiescence is surrounded by the accretion 
disk whose 
inner part is the geometrically thick and optically thin advection 
dominated accretion flow (ADAF) and the outer part is the geometrically 
thin and optically thick disk which becomes thermally unstable when 
the disk becomes too hot for hydrogen to remain neutral (Narayan, Yi 1995; 
Shakura, Sunyaev 1973; Osaki 1974).  This model of the outburst mechanism 
is called the disk instability model and satisfactorily explains the 
outburst cycle of a few tens of years and the outburst duration by 
the viscous diffusion time scale of the accretion disk in the cool and hot 
states, respectively (Mineshige 1996).

Almost all observed SXTs are distributed on the galactic disk and few 
had been discovered in the galactic halo (Chen et al. 1997; Bradt et al. 
2000).  White, van Paradijs (1996) have studied the galactic distribution 
of BHCs low-mass X-ray binaries and found an rms value for the distance 
from the galactic disk to be $\sim 0.4$ kpc.  We can easily understand 
these biased distribution because massive stars, which are responsible 
for producing neutron stars and black holes, have been generated in the 
disk rather than the halo.  

A new SXT, XTE J1118+480 whose galactic latitude is high ($\sim 62^\circ$), 
has been discovered at an intensity of 39 mCrab with the 
All-Sky Monitor (ASM; Levine et al. 1996) on the {\it Rossi X-Ray Timing 
Explorer} (RXTE) in 2000 March 29 (Remillard et al. 2000).  The X-ray 
spectrum just after the discovery is similar to Cyg X-1 in its hard 
state (Remillard et al. 2000).  The hard X-ray spectrum is well 
characterized by a power law with a photon index of 2.1 and the source 
is visible up to 120 keV which implies that it is a possible 
black hole X-ray transient (Wilson, McCollough 2000).    
In this paper, we report the discovery of the optical 
counterpart of XTE J1118+480 and the optical short-time variability, 
and discuss the binary nature and a distance estimation.

\section{Discovery and Observations}
%Figure 1
\begin{figure}
\centerline{
\epsfysize=4cm
\epsfbox{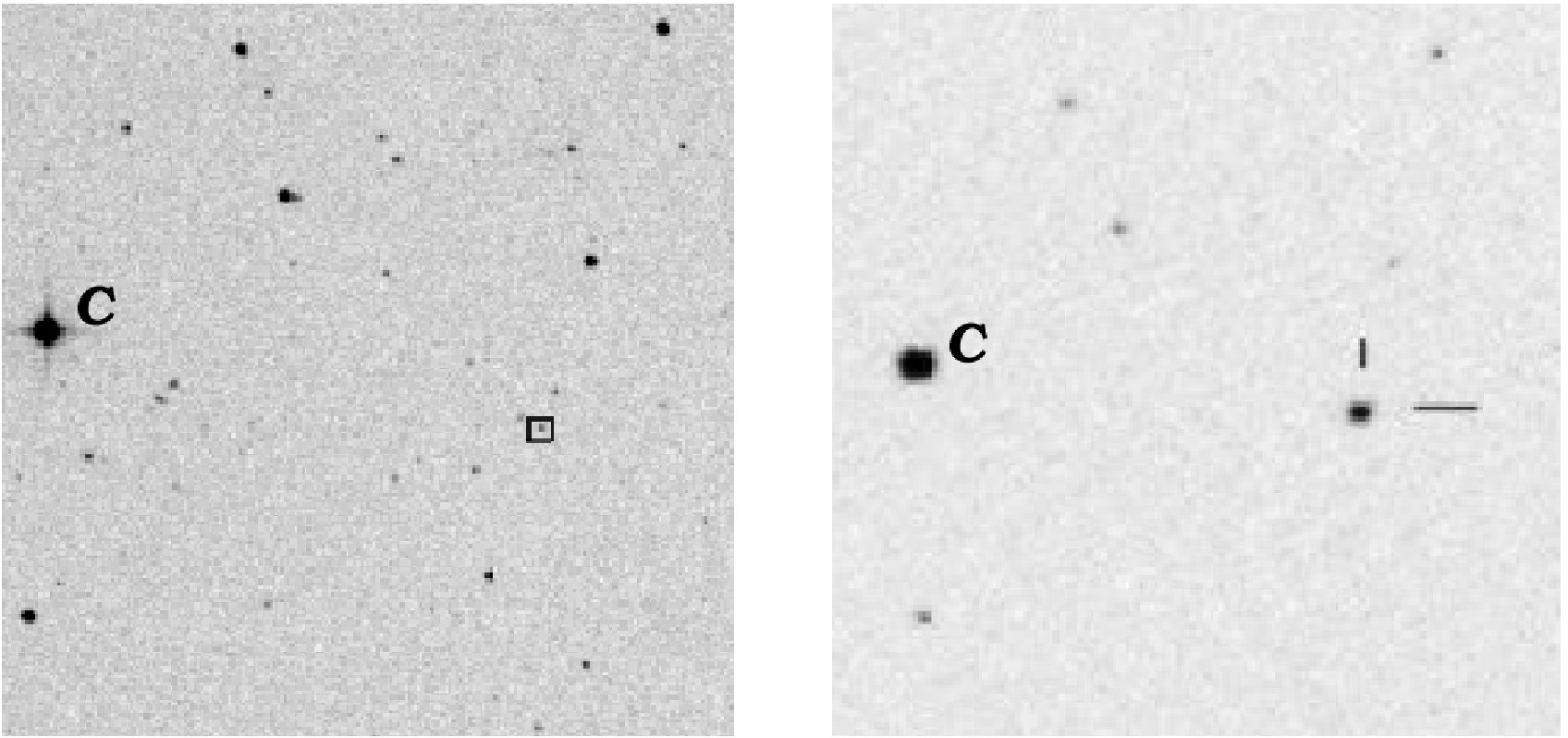}}
\end{figure}
\begin{fv}{1}
{0pc}
{CCD images of XTE J1118+480.  
The position of XTE J1118+480 is indicated.  The differential photometry 
was performed with the comparison star C.  
The left and right panels show XTE J1118+480 in quiescence (DSS image) and 
outburst (HJD 2451634.41), respectively.}
\end{fv}
Following communication of the X-ray discovery of XTE J1118+480 to the 
Variable Star NETwork (VSNET, {\tt http://www.kusastro.kyoto-u.ac.jp/vsnet}) 
on 2000 March 30, we started CCD time-series observations, which immediately 
revealed the presence of the optical counterpart at 12.92 mag, within 
the error circle of $6^\prime$ proposed by RXTE/ASM (Uemura et al. 2000a). 
The position of the optical counterpart is 
R.A. = $11^{\rm h}\,18^{\rm m}\,10^{\rm s}.85$, 
Decl. = $+48^\circ\,02^\prime\,12^{\prime \prime}.9$ 
(equinox 2000.0; accuracy $0.2^{\prime \prime}$).  
A star of 18.8 mag in the USNO A1.0 and A2.0 catalogues (USNO A1.0 
1350.08089912 = USNO A2.0 1350.07924726) lies within $2^{\prime \prime}$ 
of this position.  No other object can be seen on images of the second 
DSS images available from the Space Telescope Science Institute within 
$10^{\prime \prime}$, indicating that this star is the quiescent optical 
counterpart.  
%Figure 2
\begin{figure}
\centerline{
\epsfysize=5.5cm
\epsfbox{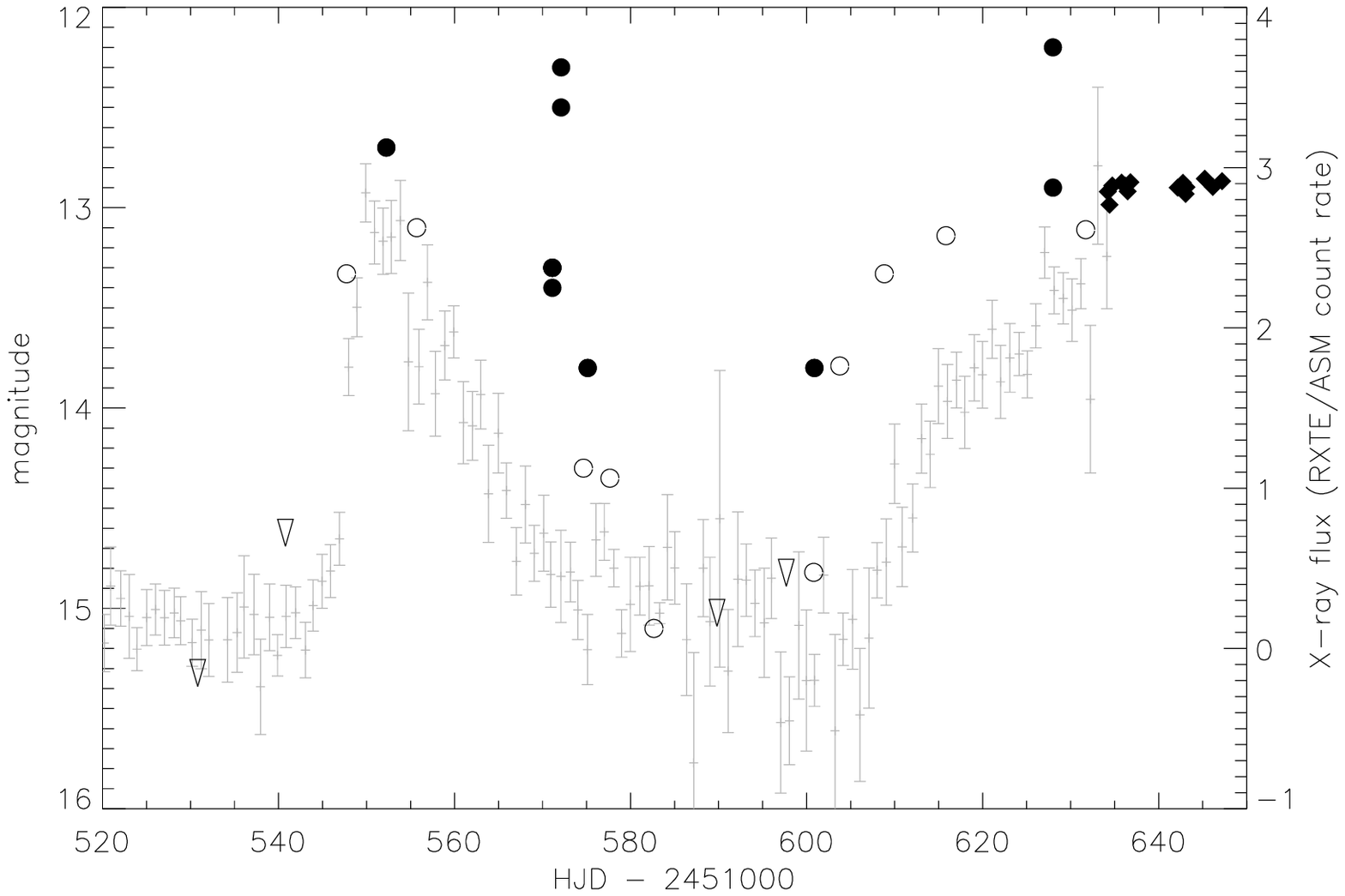}}
\end{figure}
\begin{fv}{2}
{0pc}
{Light curve of XTE J1118+480 in optical and X-ray, including prediscovery 
observations.  The abscissa denotes time in heliocentric julian date.  
The filled circles, filled diamonds, open circles, open triangle, and 
gray crosses with errorbars denote the photographic magnitude, 
nightly averaged unfiltered CCD magnitude, the $V$-magnitude and upper-
limit reported by ROTSE collaboration, and the X-ray flux observed with 
RXTE/ASM (http://xte.mit.edu/ASM\_lc.html), respectively 
(Uemura et al. 2000a; Wren 2000; Remillard 2000).  
The average error of each photometric observation is less than 0.1 mag.  
The discovery date correspond to the first filled 
diamonds on HJD 2451634.}
\end{fv}
Figure 1 gives the finding chart of XTE J1118+480.  The left and right 
panels show the DSS image and outburst image on March 30.83 (UT), 
respectively.  After detection of March outburst in optical and X-ray 
wavelengths, the reanalysis of previous data revealed the presence of 
another X-ray outburst and simultaneous optical activity in 2000 January 
(Remillard et al. 2000; Uemura et al. 2000a; Wren, McKay 2000).

%Figure 1

The CCD photometric observations were made using the following cameras and 
telescopes; unfiltered ST-7 and 25-cm Schmidt-Cassegrain (Kyoto), 
unfiltered SXL8 and 33-cm Newtonian (Conder Brow Observatory), CB245 
with clear filter and 44-cm Newtonian (California), unfiltered ST-7 
and 28-cm Schmidt-Cassegrain (Ceccano), and unfiltered ST-9e and 28-cm 
Schmidt-Cassegrain (Nayoro), whose integration times are $30\,{\rm s}$ 
, $30\,{\rm s}$, $16\,{\rm s}$, $40\,{\rm s}$, and $25\,{\rm s}$, 
respectively.  After correcting for the standard de-biasing and flat 
fielding, we processed object frames with the PSF and aperture photometry 
packages.  We performed differential photometry relative to the comparison 
star, C, shown in figure 1, whose constancy was confirmed by check star of 
GSC3451.938.  Photographic observations were performed using an unfiltered 
T-Max400 film with 10-cm camera at Nagano and Aichi.

\section{Results}

We observed no activity brighter than 15 mag from HJD 2449660 
(1994 November) to 2451547 (1999 December) with 17 negative observations.  
Figure 2 shows the light curve of XTE J1118+480 after it became active in 
2000 January.  The abscissa denotes time in heliocentric julian date.  
The filled circles, filled diamonds, open circles, and open triangle denote 
the photographic magnitude, nightly averaged unfiltered CCD magnitude 
which almost equals to $R_c$-magnitude, the $V$-magnitude, and 
upper-limit reported by ROTSE collaboration, respectively (Uemura et al. 2000a; 
Wren, McKay 2000).  The error of each photometric observation is typically 
less than 0.1 mag.  We also depict the X-ray flux observed with 
RXTE/ASM shown in gray crosses with errorbars 
(http://xte.mit.edu/ASM\_lc.html).  The discovery date 
corresponds to the first of the filled diamonds on HJD 2451634.  
%Figure 3
\begin{figure}
\centerline{
\epsfysize=8.5cm
\epsfbox{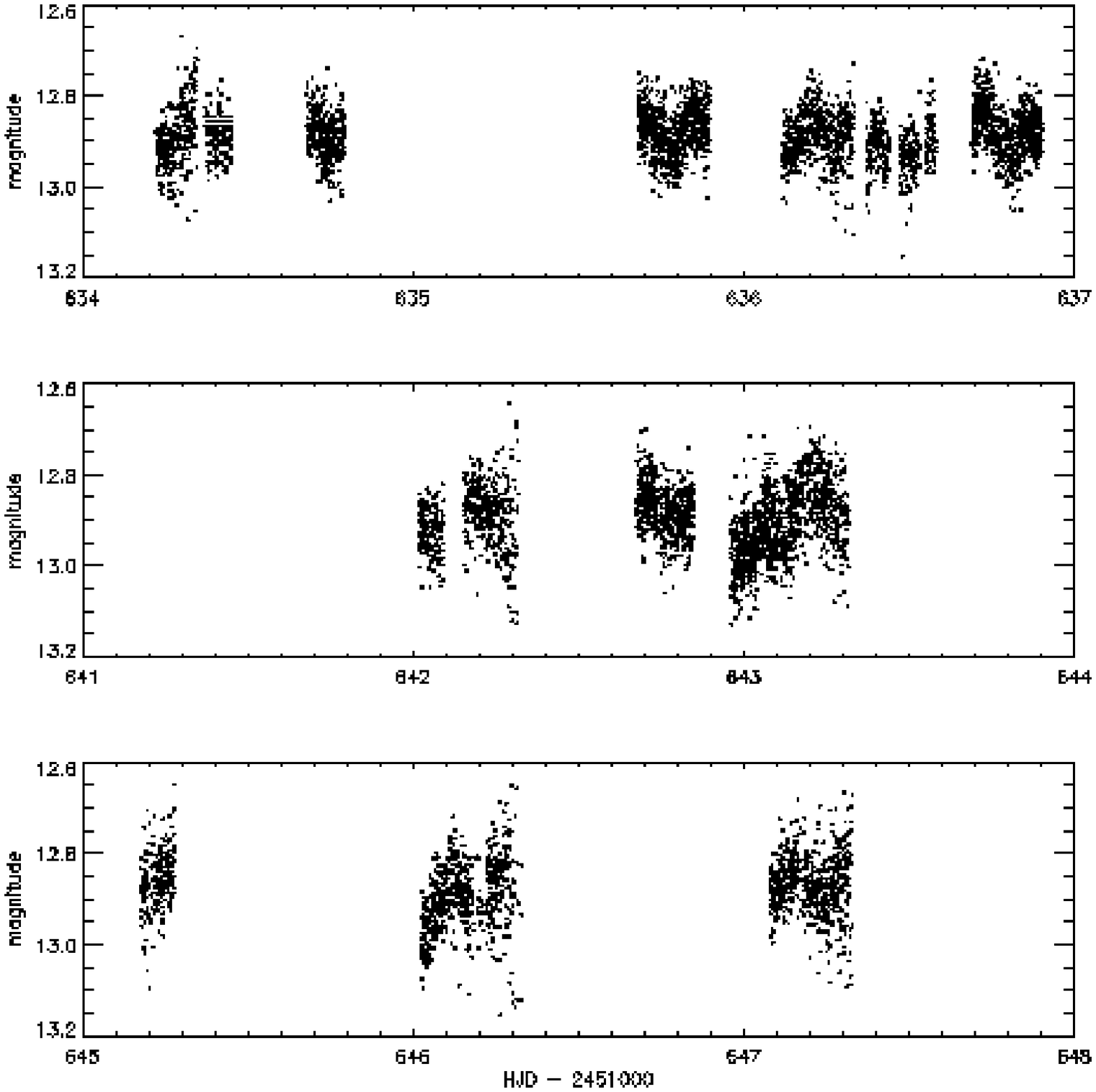}}
\end{figure}
\begin{fv}{3}
{0pc}
{Light curves of XTE J1118+480 showing short time modulation.  
The abscissa and the ordinate denote the time in HJD and magnitude.}
\end{fv}
In figure 2, we can see two distinct X-ray outbursts reported by 
Remillard et al. (2000) and two optical outbursts which correlate 
with X-ray intensity.  At the beginning of the second outburst, figure 2 
clearly shows the optical precursor which began to rise about 10 days 
prior to the X-ray outburst.  We also see a fast optical flare around 
HJD 2451572 which anti-correlates with the X-ray intensity.  The flare 
occurred just before the first X-ray outburst ended and lasted only for a 
few days.  A possible optical flare is also seen around HJD 2451627.  

%Figure 2

Figure 3 shows the light curve of XTE J1118+480 from HJD 2451634 to 
2451647.  In figure 3, we can see the sinusoidal periodic modulation 
upon the plateau of 12.89 mag and no general tendency of rise nor decay 
was observed on the limited data available.  The data on HJD 2451643 
were special in that they showed a rise over $0.2\,{\rm d}$, which was 
not seen in the other data.  Although we can see a possible hump 
feature during this rise, we have excluded this segment from the 
following period analysis.  We performed a period analysis using the 
PDM method (Stellingwerf 1978).  
%Figure 4
\begin{figure}
\centerline{
\epsfysize=5.5cm
\epsfbox{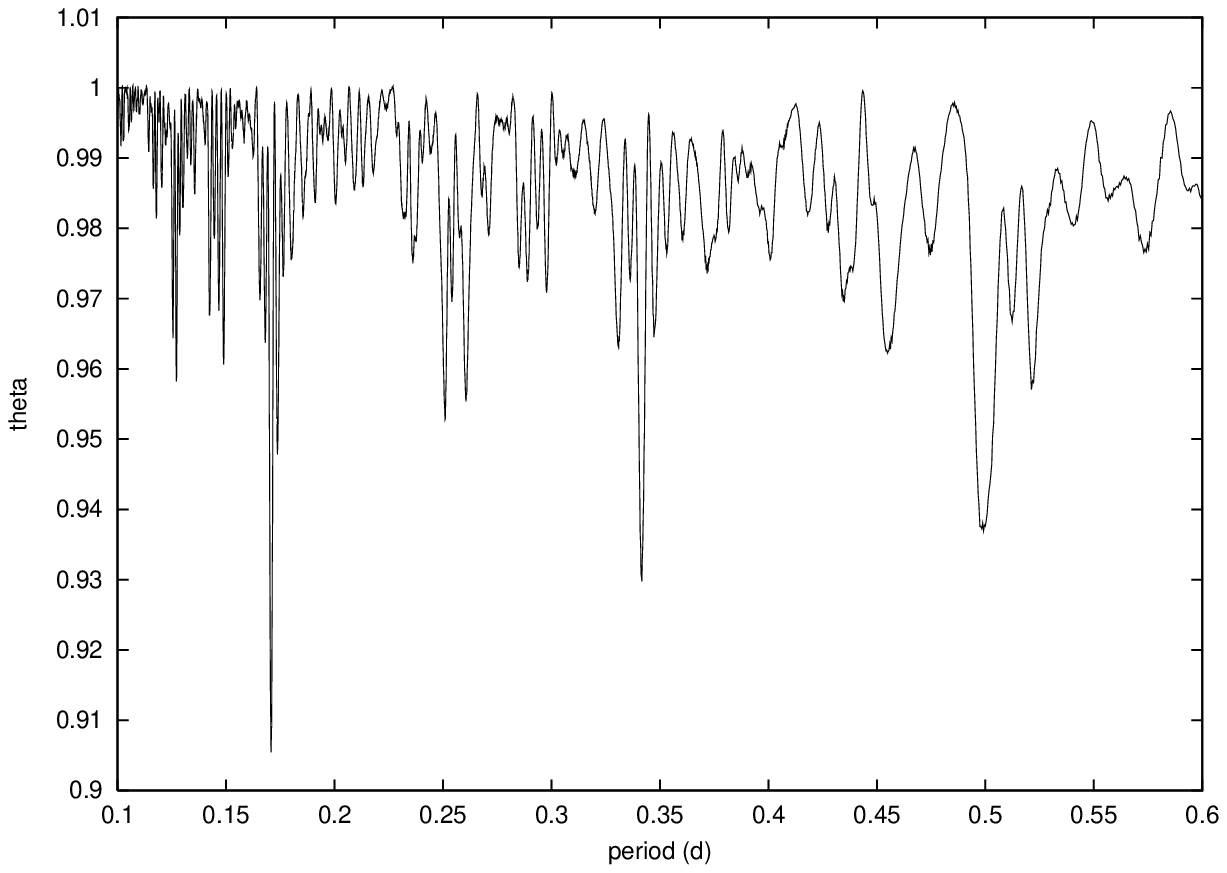}}
\end{figure}
\begin{fv}{4}
{0pc}
{Period-theta diagram calculated with PDM method using the data in 
figure 3 except for that on HJD 2451643.  The best candidate of the 
period is $0.17078\pm 0.00004$ d.}
\end{fv}
The resultant period-theta diagram 
is shown in figure 4.  As can be seen in figure 4, the best candidate 
of the period is $0.17078\pm 0.00004\,{\rm d}$.  This period is also 
confirmed with the subsequent observations by VSNET collaboration team 
(Uemura et al. 2000b).  We see a remarkable dip around
HJD 2451643.0 while the all other data well fits with the single peak 
averaged light curve.  The amplitude of 0.055 mag is not changed during 
our time-series CCD observations.  We noted that the scatter seemed 
large for a star of this brightness even if subtracting the periodic 
modulation, indicating the presence of fluctuations on a short (seconds) 
time scale.  

%Figure 3
%Figure 4

\section{Discussion}

XTE J1118+480 shows some features different from those of typical X-ray 
transients, that is, the multiple peaks within three months observed both 
in X-ray and optical, an extremely low optical to X-ray flux ratio, and 
the high galactic latitude of $62^\circ$.  For the first outburst shown 
in figure 3, the light curve profile around the peak is relatively sharp 
compared with the second one and no clear time lag is seen between the 
X-ray and optical outburst.  We can interpret this correlation through 
the occurrence of an ``inside-out'' type outburst in accretion disk 
which will generate the rapid rise in X-ray and simultaneous optical 
brightening, however this picture cannot predict the fast optical flare 
at the end of outburst (Smak 1984).  On the other hand, we suggest the 
``outside-in'' type outburst for the second outburst to explain the 
optical precursor.  To observe these different types of outburst indicates 
that they may be caused by the different instability faculty such as 
normal and super outbursts of SU\,UMa-type dwarf novae (Warner 1995).  

From the X-ray flux of 39 mCrab reported in Remillard et al. (2000) 
and the simultaneous magnitude of optical counterpart of $\sim$ 13 mag, 
the optical to X-ray flux ratio is calculated as $\sim 5$ 
which is by two orders of magnitude smaller than the typical value of 
$\sim 500$.  The X-ray intensity particularly shows a low value rather 
than optical, and this suggests an idea that it may be a high 
inclination system (Garcia et al. 2000).  The HST observation 
revealed the continuum slope somewhat flatter, which suggests the 
intrinsic X-ray flux is relatively low and hence removes the suggestion 
of the high inclination system (Haswell et al. 2000).  A high 
inclination system generally provides a high probability of occurrence 
of eclipses by secondary, whereas we have not detected them.  

It is almost certain that the periodicity appeared in figure 3 reflects 
the orbital motion of the underlying binary, through the reprocessing the
X-rays by the secondary, or possibly superhump modulations as seen
in SU\,UMa-type dwarf novae (O'Donoghue, Charles 1996).  We therefore 
suggest the orbital period of $0.17078$ d, which is 
relatively short compared with other low mass X-ray binaries 
(Ritter, Kolb 1998).  
 
Since the quiescent magnitude is 18.8 mag, if we assume a K dwarf 
secondary star and no other optical source in quiescence, 
the distance is at least 1.5 kpc, that is, 1.3 kpc 
above the galactic plane.  The color of the quiescent counterpart 
in the USNO catalog ($b-r=+0.6$) is much bluer than that of a K or M 
dwarf, implying the substatial contribution from the accretion disk.  
The above distance estimate should be thus regarded as a lower limit, 
implying that XTE J1118+480 is a galactic halo X-ray transient.  
We can generally expect that such a galactic halo object is older than 
that in disk, and this is consistent with the short orbital period 
of 0.17078 d which implies a more evolved binary system.  It should be 
noted, however, if we assume a short orbital period of 0.17078 d, the 
secondary may be a smaller star than in other typical SXTs.  This case 
may be comparable to the system GRO J0422+32, whose orbital parameter 
and secondary type are 0.212 d and M2V, respectively (Orosz, Bailyn 1995; 
Filippenko et al. 1995).  An M2V star secondary leads to the 
lower limit of the distance of 0.55 kpc.  Optical or infrared spectroscopic 
observations are definitely needed to unambiguously determine the orbital 
period and the type of the secondary.

On the point of the high galactic latitude SXTs, 3U 0042+32 is similar 
to XTE J1118+480 (Ricketts, Cooke 1977).  The outburst in 1977 of 
3U 0042+32 was detected by {\it Uhuru} which classified it as a high 
galactic latitude ($\sim -30^\circ$) X-ray transient.  During the 
active phase in 1977 Feburuary, 3U 0042+32 experienced at least four 
distinct outbursts whose interval is estimated as 11.6 d (Watson, 
Ricketts 1978).  Compared with XTE J1118+480, the outburst duration 
seems to be quite shorter (decay e-folding time $= 2.8$ d).  Another 
noteworthy, possibly related, high galactic latitude X-ray binary
is MS 1603+2600 = UW CrB (Morris et al. 1990).  This object is a persistent
source whose orbital period is 111 min.  Although Hakala et al. (1998)
suggested this X-ray binary may correspond to a quiescent state of SXTs,
the discovery of a short-period transient system XTE J1118+480 may suggest
that these two systems comprise a new variety of halo X-ray binaries.

\section{Summary}

We discovered the optical counterpart of the high galactic latitude 
soft X-ray transient, XTE J1118+480, whose position is 
R.A. = $11^{\rm h}\,18^{\rm m}\,10^{\rm s}.85$, 
Decl. = $+48^\circ\,02^\prime\,12^{\prime \prime}.9$.  
There are 18.8 mag star within the $2^{\prime \prime}$ error circle in USNO 
catalogue and we consider it as an optical counterpart in quiescence.  
We revealed the two distinct optical outbursts before 
discovery.  From our time-series CCD photometry, we detected a 
0.17078 d periodicity which we consider as the orbital period.  
If we assume the secondary star of K dwarf, XTE J1118+480 is the 
possible first firmly identified black hole candidate X-ray transient 
in the galactic halo.

\section*{References}
\re
Bailyn C. D., Jain R. K., Coppi P., Orosz J. A.\ 1998, ApJ\ 499, 367
\re
Bradt H., Levine A., Remillard R., Donald A. S.\ astro-ph/0003438
\re
Chen W., Shrader C. R., Livio M.\ 1997, ApJ\ 491, 312
\re
Filippenko A. V., Matheson T., Ho L. C.\ 1995, ApJ\ 455, 614
\re
Garcia M., Brown W., Pahre M., McClintock J.\ 2000, IAU Circ. 7392
\re
Hakala P. J., Chaytor D. H., Vihu O., Piirola V., Morris S. L., Muhli P.\ 1998, ApJ\ 333, 540
\re
Haswel C. A., Hynes R. I., King A. R.\ 2000, IAU Circ. 7407
\re
Levine A. M., Bradt H., Cui W., Jernigan J. G., Morgan E. H., Remillard R., Shirey R. E., Smith D. A.\ 1996, ApJ\ 469, L33
\re
Mineshige S.\ 1996, PASJ\ 48, 93
\re
Morris S. L., Liebert J., Stocke J. T., Gioia I. M., Schild R. E., Wolter A.\ 1990, ApJ\ 365, 686
\re
Narayan R., Yi I.\ 1995, ApJ\ 452, 710
\re
O'Donoghue D., Charles P. A.\ 1996, MNRAS.\ 282, 191
\re
Orosz J. A., Bailyn C. D.\ 1995 ApJ\ 446, L59
\re
Orosz J. A., Jain R. K., Bailyn C. D., McClintock J. E., Remillard R. A.\ 1998, ApJ\ 499, 375
\re
Osaki Y.\ 1974, PASJ\ 26, 429
\re
Remillard R., Morgan E., Smith D., Smith E.\ 2000, IAU Circ. 7389
\re
Ricketts M. J., Cooke B. A.\ 1977, IAU Circ. 3039
\re
Ritter H., Kolb U.\ 1998, A\&AS\ 129, 83
\re
Shakura N. I., Sunyaev R. A.\ 1973, A\&A\ 24, 337
\re 
Smak J.\ 1984, Acta Astron. 34, 161
\re
Stellingwerf R. F.\ 1978, ApJ\ 224, 953
\re
Tanaka Y., Shibazaki N.\ 1996, ARA\&A\ 34, 607
\re
Uemura M., Kato T., Yamaoka H.\ 2000a, IAU Circ. 7390
\re
Uemura M., Kato T., Matsumoto K., Honkawa M., Cook L. M., Martin B., 
Masi G., Oksanen A., Moilanen M., Novak R., Sano Y., Ueda Y.\ 2000b, IAU Circ. 
7418
\re
van Paradijs J., McClintock J. E.\ 1995, in X-ray Binaries, ed. W.H.G. Lewin J., van Paradijs E. P. J. van den Heuvel (Cambridge: Cambridge Univ. Press), 58
\re
White N., van Paradijs J.\ 1996, ApJ\ 473L, 25
\re
Warner B. \ 1995, Cataclysmic Variable Stars, p126--215 (Cambridge Univ. Press, Cambridge)
\re
Watson M. G., Ricketts M. J.\ 1978, MNRAS.\ 183, 35P
\re
Wilson C. A., McClollough M. L.\ 2000, IAU Circ. 7390
\re
Wren J., McKay T.\ 2000, IAU Circ. 7394

\end{document}